\documentclass[journal]{IEEEtran}
\usepackage{graphicx}
\usepackage{amssymb}
\usepackage{array}
\usepackage{multirow}

\begin{document}
\title{Performances and ageing study of resistive-anodes Micromegas detectors for HL-LHC environment}

\author{F.~Jeanneau$^{*,a}$, T.~Alexopoulos$^b$, D.~Atti\'e$^a$, M.~Boyer$^a$, J.~Derr\'e$^a$, G.~Fanourakis$^c$, E.~Ferrer-Ribas$^a$, J.~Gal\'an$^a$, E.~Gazis$^b$, T.~Geralis$^c$, A.~Giganon$^a$, I.~Giomataris$^a$, S.~Herlant$^a$, J.~Manjarr\'es$^a$, E.~Ntomari$^c$, Ph.~Schune$^a$, M.~Titov$^a$, G.~Tsipolitis$^b$
\thanks{Manuscript received November 14, 2011.}%
\thanks{$^*$ Corresponding author (fabien.jeanneau@cea.fr, tel: +33(0)169083965).}%
\thanks{$^a$ IRFU, CEA-Saclay, 91191 Gif-sur-Yvette, France.}%
\thanks{$^b$ National Technical University of Athens, Athens, Greece.}%
\thanks{$^c$ Institute of Nuclear Physics, NSCSR Demokritos Athens, Greece.}%
}

\maketitle
\pagestyle{empty}
\thispagestyle{empty}

\begin{abstract}
With the tenfold luminosity increase envisaged at the HL-LHC, the background (photons, neutrons, $\bf ...$) and the event pile-up probability are expected to increase in proportion in the different experiments, especially in the forward regions like, for instance, the muons chambers of the ATLAS detector.
Detectors based on the Micromegas \cite{micromegas} principle should be good alternatives for the detector upgrade in the HL-LHC framework because of a good spatial ($\bf <100\mu$m) and time (few ns \cite{time_res}) resolutions, high-rate capability, radiation hardness, good robustness and the possibility to build large areas.
The aim of this study is to demonstrate that it is possible to reduce the discharge probability and protect the electronics by using a resistive anode plane in a high flux hadrons environment. Several prototypes of 10x10 cm$\bf ^2$, with different pitches (0.5 to 2 mm) and different resistive layers have been tested at CERN ($\bf \pi^+$@SPS).  
Several tests have been performed with a telescope at different voltages to assess the performances of the detectors in terms of position resolution and efficiency. The spark behaviour in these conditions has also been evaluated. 
Resistive coating has been shown to be a successful method to reduce the effect of sparks on the efficiency of micromegas. A good spatial resolution ($\bf \sim 80\mu$m) can be reached with a resistive strip coating detector of 1mm pitch and a high efficiency ($\bf > 98\%$) can be achieved with resistive-anode micromegas detector. 
An X-rays irradiation has been also performed, showing no ageing effect after more than 21 days exposure and an integrated charge of almost 1C. 
\end{abstract}

\section{Introduction}

\begin{figure}[htbp]
\centering
\includegraphics[width=8cm]{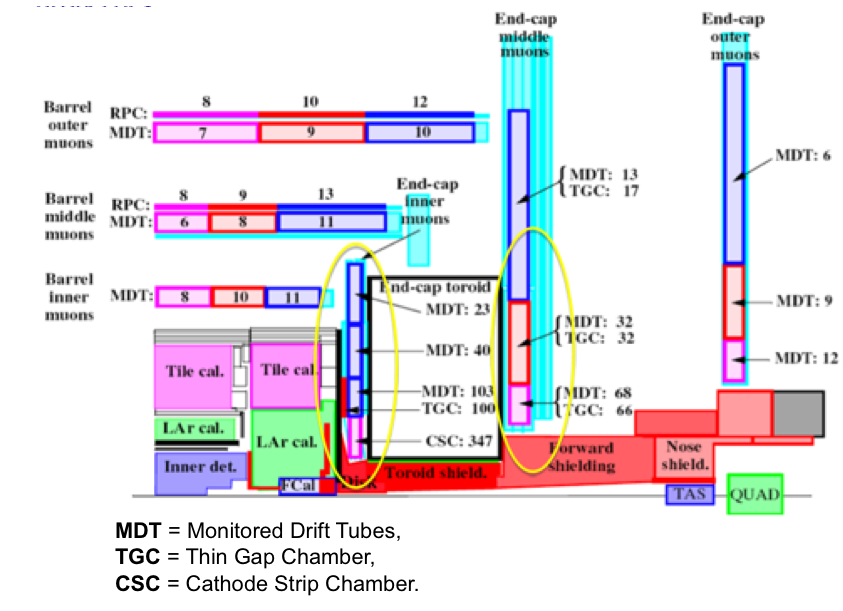}
\caption{View of the Atlas detector \cite{atlas_bkg}. The detectors in the highest rapidity region will need to be upgraded.}
\label{atlas_section}
\end{figure}

\IEEEPARstart{W}{ith} the Large Hadron Collider (LHC) running and ramping in energy and luminosity, plans are already advancing for an upgrade. The High Luminosity LHC (HL-LHC) project, the luminosity upgrade is expected to be in two stages: first by a factor of three in collision rate -- $3\cdot10^{34} cm^{-2}s^{-1}$, for phase-I; then by a factor of ten for the nominal HL-LHC phase-II \cite{atlas_upgrade}. The particularly harsh background environment in the detectors at the HL-LHC places a number of severe constraints on the performance of such detectors. Counting rates will grow up to 20 kHz/cm$^2$ in the most unfavourable regions of the ATLAS muon system \cite{atlas_bkg}. To cope with the corresponding increase in background rates, ATLAS experiment muon system will likely need major changes, at least in the highest rapidity region, see Fig. \ref{atlas_section}. Based on background estimations at HL-LHC situation a list of requirements for these new detectors has been established:
\begin{itemize}
\item High counting rate capability, including dense ionization;
\item	High single plane detection efficiency ($\ge98\%$);
\item	Spatial resolution better than 100 $\mu$m, possibly up to large incident angles (45$^\circ$);
\item Second coordinate measurement with a few mm precision;
\item Two-track discrimination at a distance of 1-2 mm.
\end{itemize}

The MAMMA collaboration \cite{mamma}, started at CERN since 2007, propose Micromegas as a good candidate to equip the future Atlas small wheel for the HL-LHC.

\section{Resistive-anodes Micromegas detectors}

\begin{figure}[htbp]
\centering
\includegraphics[width=8cm]{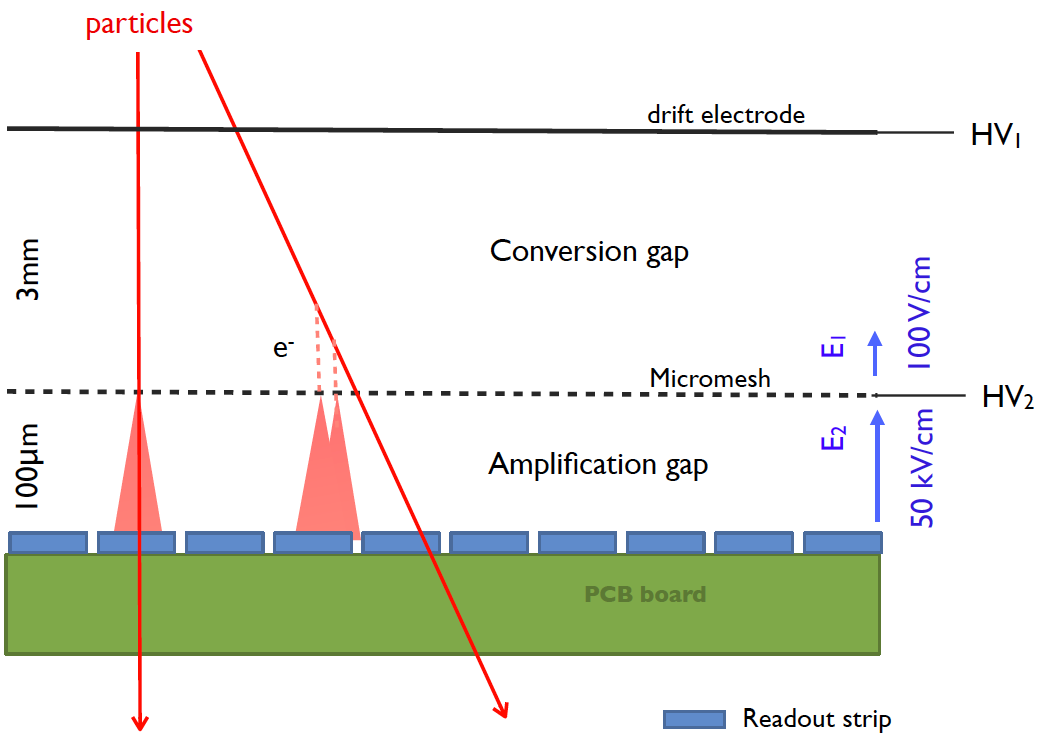}
\caption{Principle of a Micromegas detector. Sparks can occur between the strips and the micromesh}
\label{mM_principle}
\end{figure}

In standard Micromegas detector, a very high amplication field is applied in a very thin gap, as it is represented on the Fig. \ref{mM_principle}. In HL-LHC environment, the high flux of hadrons can produce highly ionizing events that leads to large energy deposit and an increasing probability for sparks occurance \cite{sparks}.

Sparks are not a problem concerning the detector robustness or the electronics on which we can had a suitable protection \cite{protec}, but the discharge of the whole micromesh induces a dead time up to 1 or 2 ms to recover the nominal voltage. One of the suggested solutions to reduce the effect of sparks is to use resistive coatings on top of the read-out strips. This technology was first used for the readout of Time Projection Chambers to spread out the signal of the avalanche and reduce the number of electronic channels \cite{dixit}. The micromegas structure is then built on top of the resistive strips or layer. During this test we studied three different resistive-anodes and geometries configurations:\\

\begin{figure}[htbp]
\centering
\includegraphics[width=7cm]{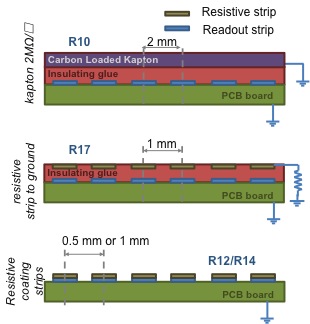}
\caption{Geometry of the different resistive-anodes technologies}
\label{resistive}
\end{figure}

\begin{itemize}
\item {\bf Carbon-Loaded Kapton ($2\, \mathrm{M}\Omega/ \square$):} The plane of strips is covered with an insulating layer of 75 $\mu$m thickness, on which is glued a foil of carbon-loaded kapton ($2\, \mathrm{M}\Omega/ \square$ resistive layer), see Fig. \ref{resistive} -- {\bf R10}.\\

\item	{\bf Resistive strip to ground:} The plane of strips is covered with an insulating layer of 64 $\mu$m thickness, on which resitive strips, matching the geometry of the copper strips, are deposited using a resistive ink of  $100\, \mathrm{k}\Omega/ \square$. Each strip is grounded through a resistor of $30\, \mathrm{M}\Omega$ and the resistor along the strips is $250\, \mathrm{M}\Omega$, see Fig. \ref{resistive} -- {\bf R17}.\\

\item	{\bf Resistive strips:} On each copper-strip anode is deposited a resistive coating (resistive ink of  $100\, \mathrm{k}\Omega/ \square$). The resistor along the strips is $300\, \mathrm{k}\Omega$, see Fig. \ref{resistive} -- {\bf R12/R14}.
\end{itemize}

The specifications of the different technologies are summarized in the table \ref{specifs}.

 \begin{table}[htbp]
\begin{center}
\begin{tabular}{|c|c|c|c|c|}
\hline
\multirow{2}{*} {Detector} & \multirow{2}{*} {Pitch} & \multirow{2}{*} {Circuit type} & Resistor & Insulator \\
& & & measured & thickness \\
\hline
\hline
\multirow{2}{*} {R10} & \multirow{2}{*} {2.0 mm} & \multirow{2}{*} {CLK $2\, \mathrm{M}\Omega/ \square$} & \multirow{2}{*} {-} & \multirow{2}{*} {75 $\mu$m} \\
& & & & \\
\hline
\multirow{2}{*} {R17} & \multirow{2}{*} {1.0 mm} & R-strip to gnd: & Strips: $250\, \mathrm{M}\Omega$ & \multirow{2}{*} {64 $\mu$m} \\
 & & $100\, \mathrm{k}\Omega/ \square$ & To gnd: $30\, \mathrm{M}\Omega$ & \\
\hline
\multirow{2}{*}  {R14} & \multirow{2}{*} {1.0 mm} & R-coating & \multirow{2}{*} {$300\, \mathrm{k}\Omega$}  & \multirow{2}{*} {-} \\
  &  & $100\, \mathrm{k}\Omega/ \square$ & & \\
\hline
 \multirow{2}{*}  {R12} & \multirow{2}{*} {0.5 mm} & R-coating & \multirow{2}{*} {$300\, \mathrm{k}\Omega$}  & \multirow{2}{*} {-} \\
  &  & $100\, \mathrm{k}\Omega/ \square$ & & \\
\hline
\end{tabular}
\end{center}
\caption{Resistive technologies specifications.}
\label{specifs}
\end{table}

\section{Beam tests 2010}
\subsection{Setup}

\begin{figure}[htbp]
\centering
\includegraphics[width=9.5cm]{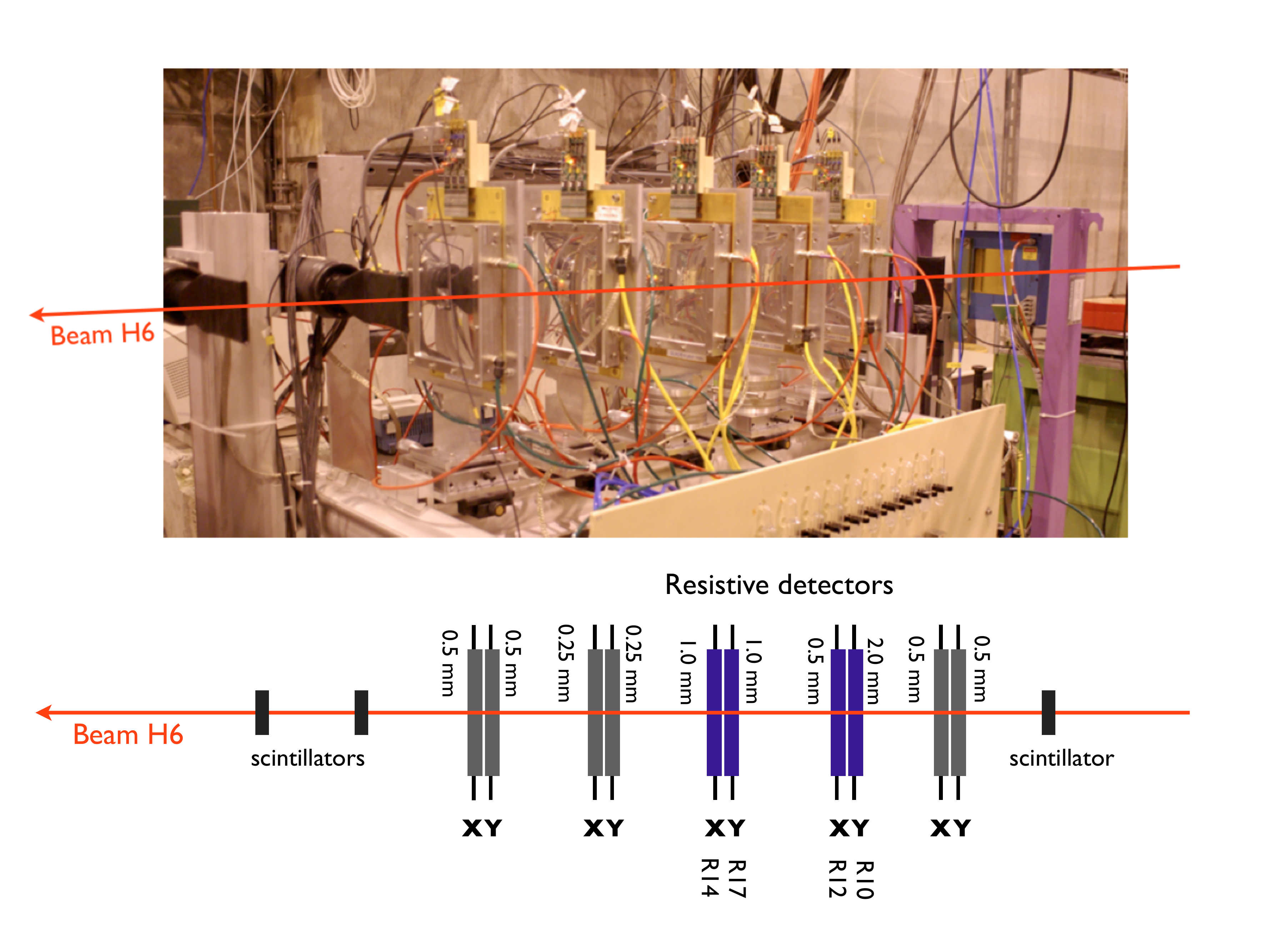}
\caption{Picture and scheme of the test beam setup. The beam direction is indicated. The telescope is made of 3 (X-Y) planes and 4 detectors are tested (10 Micromegas chambers in total).}
\label{setup}
\end{figure}

The resistive-anodes detectors described above were exposed to 120 GeV/c pions beam at CERN SPS H6 area, during two weeks in autumn of 2010. An external reference
measurement, on the plane normal to the beam direction was given by a telescope consisting of 3 (X-Y) plans of standard Micromegas. And four resistive detectors were tested in a two (X-Y)
configuration as it is represented on Fig. \ref{setup}. 

The standard detector have been manufactured at Saclay, whereas the resistive-anodes detectors were manufactured at CERN workshop. During the test beam, two different gas mixtures were used: Ar + 2\%C$_4$H$_{10}$ + 3\%CF$_4$ for the resistive chambers and Ar + 2\%C$_4$H$_{10}$ for the telescope. The resistive detectors were tested with different high voltage values and mounted on a rotating structure, in order to collect data with different beam angle.

The signal readout was performed with a GASSIPLEX \cite{gassiplex} electronics, with 96 channels multiplexed and a peaking time of 1.2 $\mu$s. The data acquisition system was based on 4 C-RAMS modules driven by a sequencer from CAEN. 

\subsection{Performances}
The goal of this tests was to evaluate the performances of the resistive-anodes detectors and check the influence of resistive material on the spatial resolution and efficiency of the detector. The details of the data analysis can be found in \cite{joany}.\\

\subsubsection{Spatial Resolution} \leavevmode\par

\begin{figure}[htbp]
\centering
\includegraphics[width=9cm]{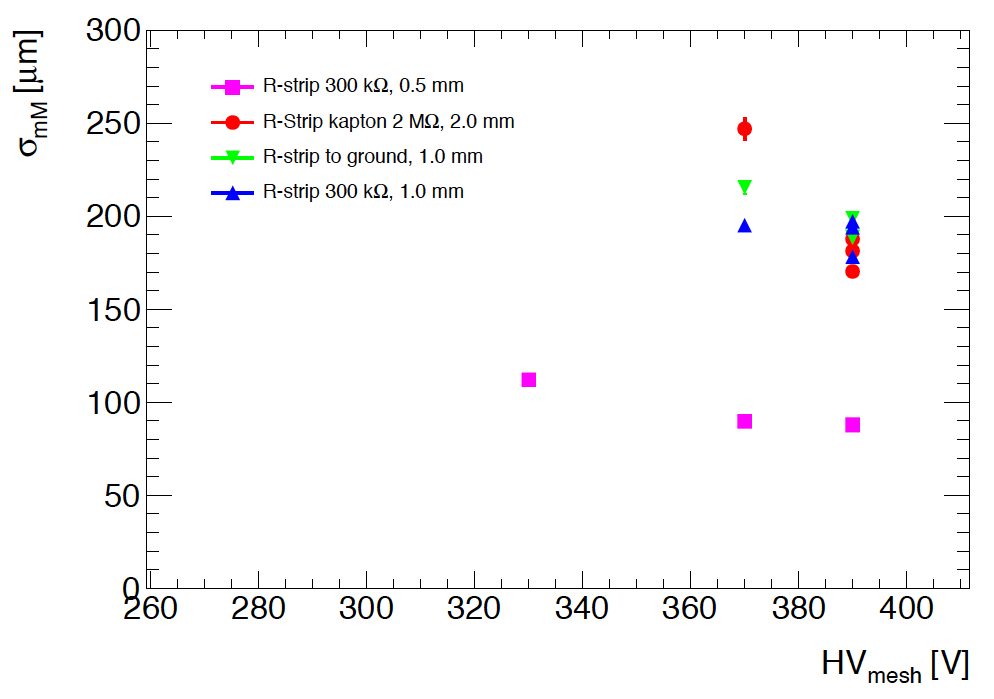}
\caption{Spatial resolution vs micro-mesh voltages for the different resistive technologies.}
\label{res_evol}
\end{figure}

The spatial resolution is calculated using reference tracks given by the telescope. The results plotted in the Fig. \ref{res_evol} are obtained by extrapolated these tracks at the level of the detector of interest. Then the distribution of the difference between the position measured and the position extrapolated is fitted by a gaussian curve of which the r.m.s. is the spatial resolution. This is clearly related to the pitch of the strips and th best resolution (88.1 $\pm 0.7 \mu$m) is obtained for a pitch of 0.5 mm.

\subsubsection{Efficiency} \leavevmode\par

\begin{figure}[htbp]
\centering
\includegraphics[width=9cm]{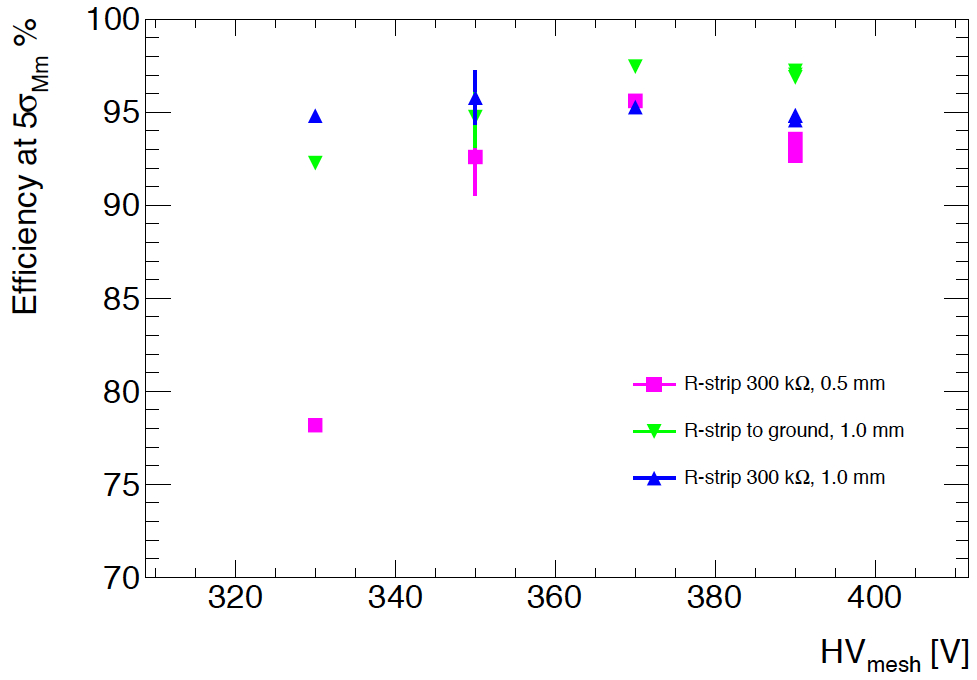}
\caption{Efficiency vs micro-mesh voltages for an acceptance window of 5 $\sigma_{mM}$ for the different resistive technologies.}
\label{eff_evol}
\end{figure}

The detector is efficient when the measured position is within a window of $\pm$5$\sigma_{mM}$ around the extrapolated position from the reference track of the telescope.
As it is shown on Fig. \ref{eff_evol}, the detection efficiency increases with the mesh voltage up to 98\%. The detector with resistive strips connected to the ground (R-strip to the ground -- R17) present the best efficiency
results.

\begin{figure}[htbp]
\centering
\includegraphics[width=9cm]{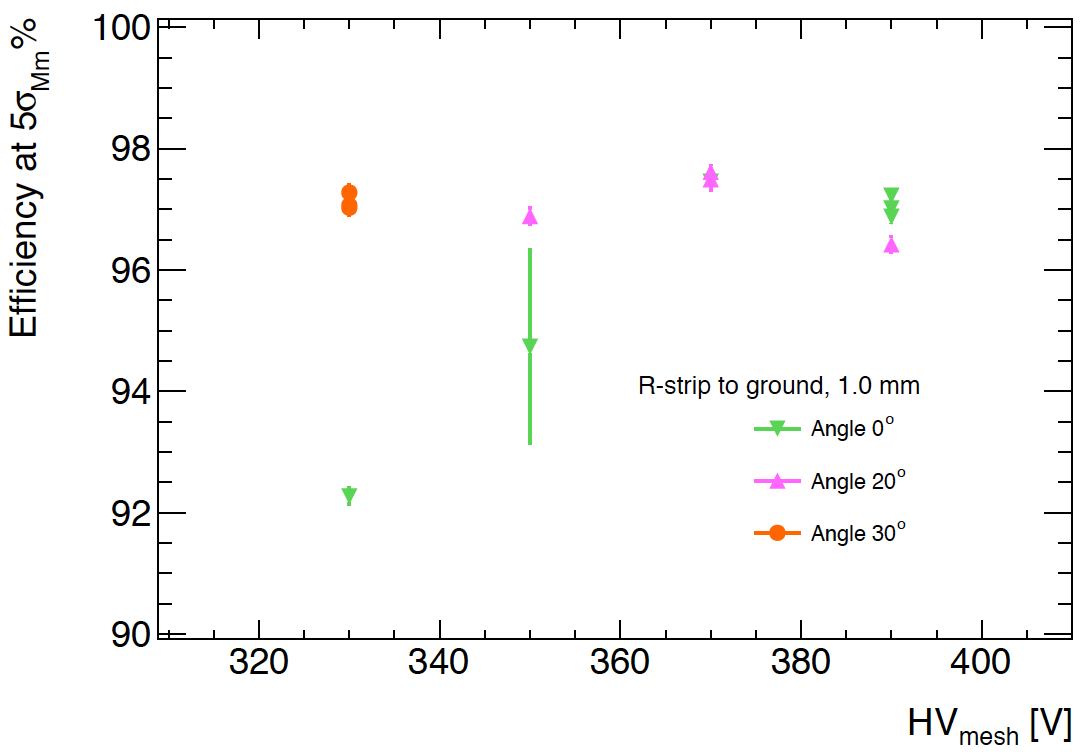}
\caption{Efficiency vs micro-mesh voltages at several incidence angles for the R-strip to ground detector.}
\label{eff_angle}
\end{figure}

In order to get tracks with different incidence angles the detectors were rotated. As the signal is spread on several strips (depending on the pitch -- see Fig. \ref{mM_principle}), we have to study the impact on the global efficiency. The Fig. \ref{eff_angle} shows the efficiency at different voltages on the micro-mesh and incidence track angle, for the R-strip to the ground detector (R17). For normal incidence tracks, the efficiency is around 98\%. For inclined tracks (the usual case in Atlas), the efficiency is of the same order and can be even better for low voltage values, because for inclined tracks the depth of gas is more important leading to more primary electron-ion pairs. Since this detector (R- to ground -- R17) has the best efficiency, it is the best candidate to be proposed for the Atlas upgrade.

\subsection{Sparks behaviour}

\begin{figure}[htbp]
\centering
\includegraphics[width=9cm]{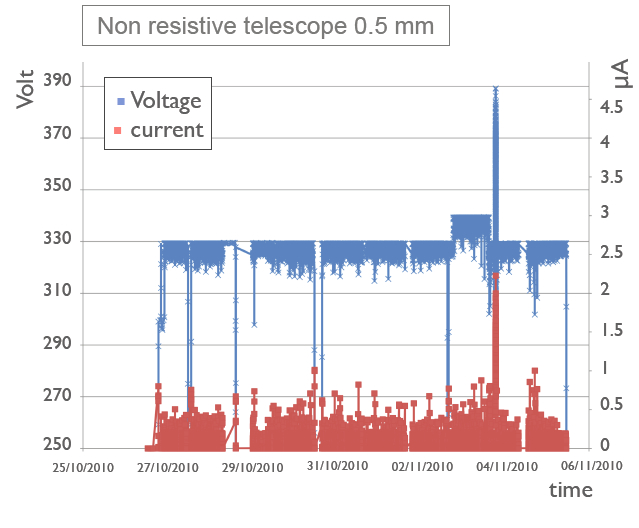} \\
\includegraphics[width=9cm]{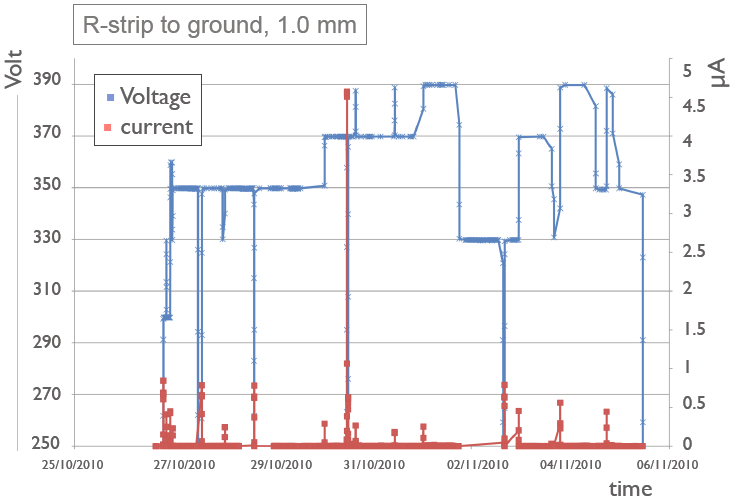} 
\caption{Micro-meshes voltages and currents monitored during ten days for a standard detector (top) and and the R-strip to ground detector (bottom).}
\label{sparks}
\end{figure}

During the ten days of the beam test, the voltages and currents were monitored by the power supply (CAEN SY2527). The results for the voltage and current on the mesh are given on Fig. \ref{sparks}. For a standard Micromegas detector -- on the left -- many current peaks and so many sparks are visible (red curve).  The current can go up to 1 or 2 $\mu$A and are related to many voltage drops (blue curve) of 10 to 20 volts, that lead to a loss of efficiency of this detector. On the right, the same curves for the R-strip to ground detector (R17) show few sparks and very low current mainly corresponding to the charging up of the micro-mesh. For this detector there is no voltage drop thus no dead time and no loss of efficiency.

\section{Ageing}

\begin{figure}[htbp]
\includegraphics[width=5.5cm]{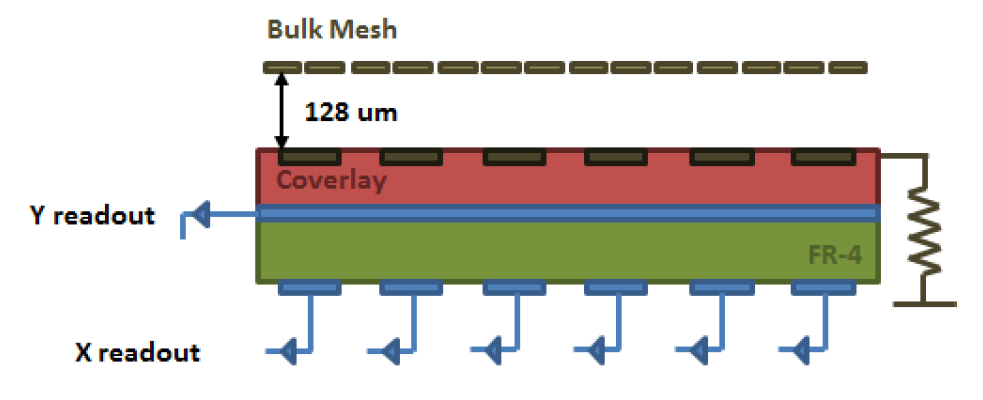} 
\includegraphics[width=3cm]{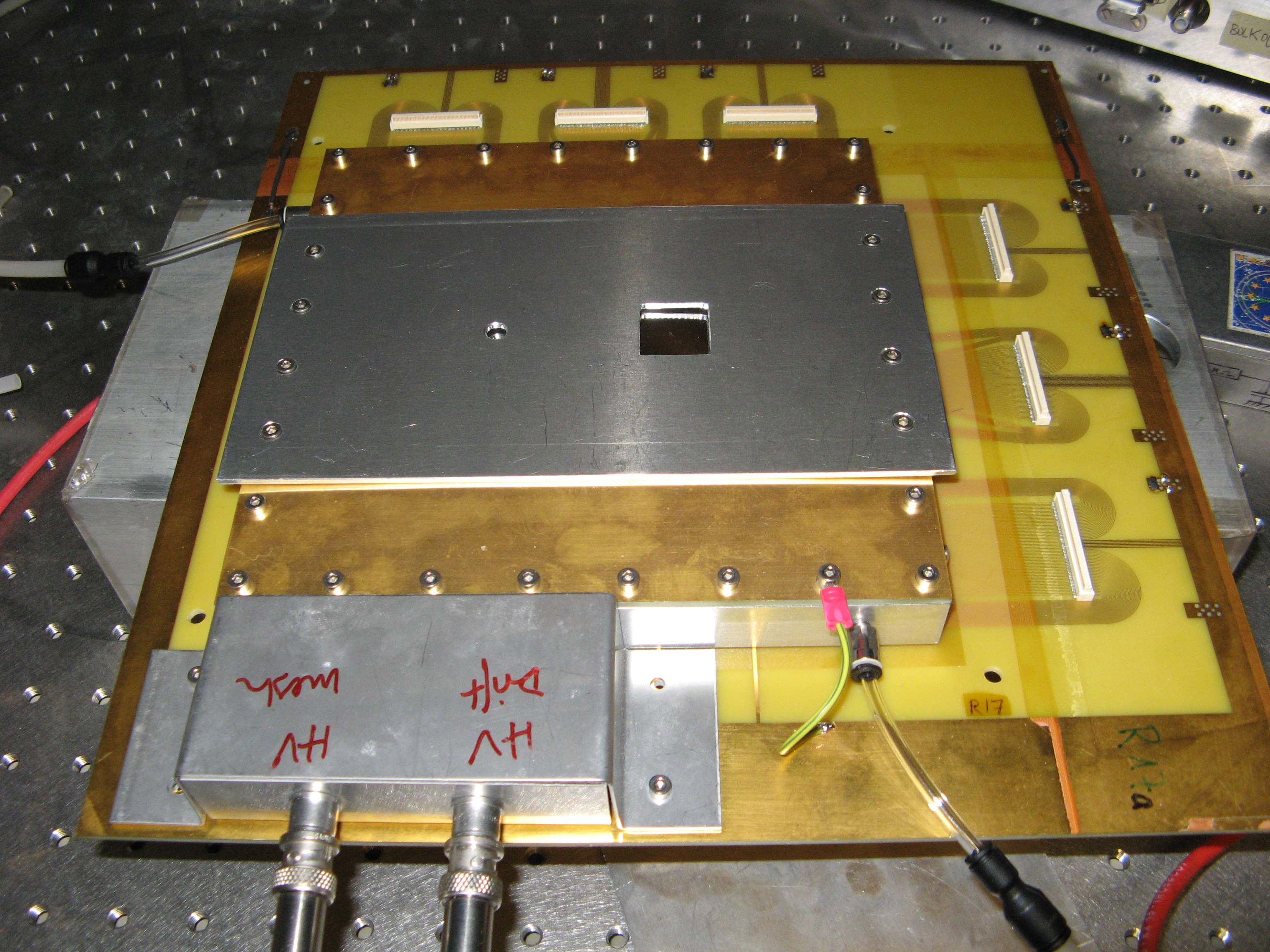} 
\caption{Section of the 2-dimensional resistive detector (left) and pictures of the detector with the metallic maks used during the irradiation (right).}
\label{ageing_det}
\end{figure}

An ageing study of resistive-anodes detector is mandatory to assess their capability to handle the rate and level of radiations at the HL-LHC. For this study, two new prototypes are used, lent by the MAMMA collaboration and built at the CERN workshop. These detectors are based on the R-strip-to-the-ground (R17) architecture but with one more layer of resistive strips, enabling a 2-dimensional readout (see Fig. \ref{ageing_det}). One prototype is kept unexposed as a reference and the other is exposed to the high rate of an X-ray  gun on an area of 4 cm$^2$ through a metallic mask as it is shown on the picture Fig. \ref{ageing_det}. 

\subsection{Integrated charge calculation}

The goal of the X-rays exposure is to accumulate a charge that is comparable to the workingtime on the HL-LHC. Taking into account the total charge $Q^{m.i.p.}$ obtained in the detector after the interaction of a minimum ionization particule in the conversion gap:

$$
Q^{m.i.p.}= \frac{\Delta E^{m.i.p.}}{W_i} \cdot q_{e^-} \cdot G
$$

where $\Delta E^{m.i.p.}$ is the m.i.p. loss of energy in 0.5 cm of an Ar+10\%CO$_2$ gas mixture (1.25 keV), $W_i$ the gas mixture ionization potential (26.7 eV), $q_{e^-}$ the electron charge and $G$ the gain of the detector (here a nominal operation gain of 5000). In these conditions $Q^{m.i.p.}=$ 34.7 fC.\\

At the HL-LHC, the expected rate in the muon chambers close to the beam pipe will be 10 kHz/cm$^2$. For 5 years of operation time (200 days/years), the total charge surface density will then be $\sigma^{HL-LHC}=$ 32,3 mC/cm$^2$.\\

The irradiated area is 4 cm$^2$, the minimum integrated charge on the X-rays gun to be equivalent at 5 years of HL-LHC will then be $\simeq$ 130 mC.

\subsection{Results}

\begin{figure}[htbp]
\centering
\includegraphics[width=9cm]{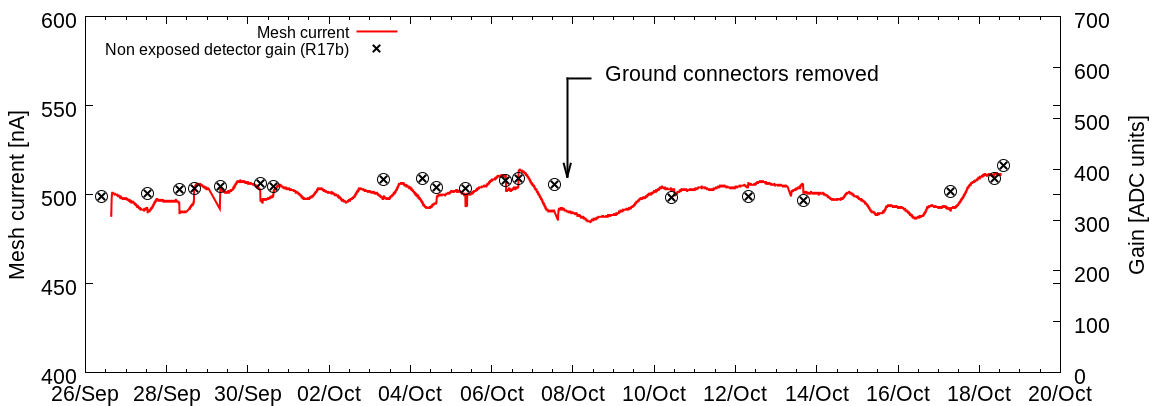} 
\caption{Mesh current evolution (red curve) for a period of 21 days and an integrated charge of 918 mC. The gain control measurements with the non exposed detector are also plotted (black circles).}
\label{current_evol}
\end{figure}

The detector has been exposed for more than 20 days with a gain of 5000 and a gas flow of one renewal per hour. The curent on the mesh has been recorded and is plotted on the Fig. \ref{current_evol}. In a first period the connectors were grounded then floating in a second period, inorder to see the influence of the intern layers on the detector behaviour. The current is very stable, and don't show any ageing effect, on the whole irradiation period which corresponds to 21.3 days of exposure and an integrated charge of 918 mC, that is 5 years of HL-LHC with a security factor more than 5.

\begin{figure}[htbp]
\includegraphics[width=9cm]{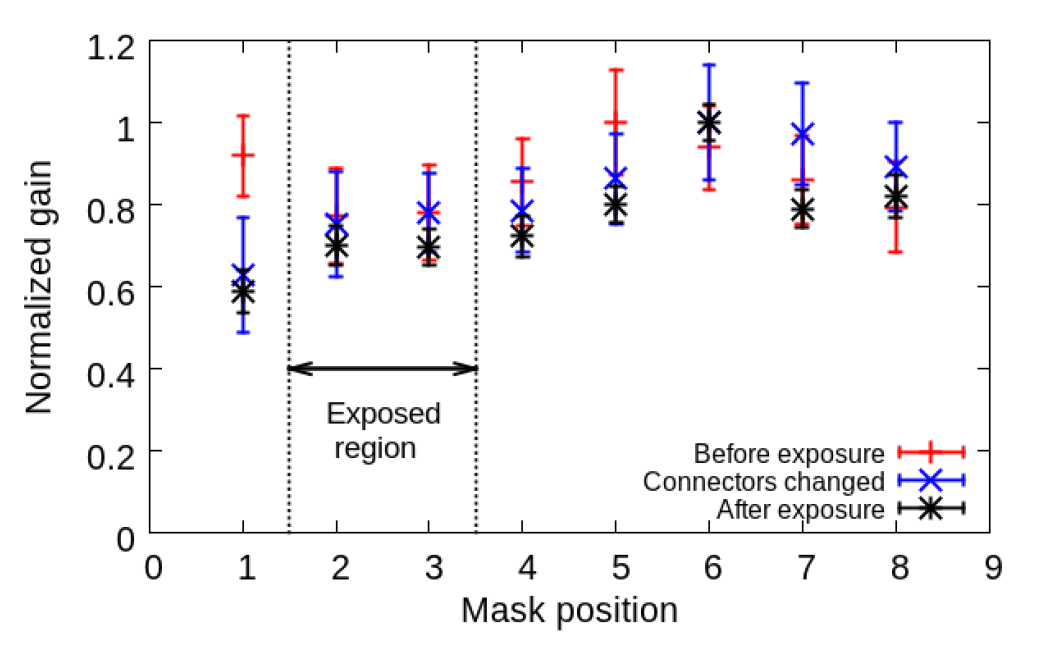} 
\includegraphics[width=9cm]{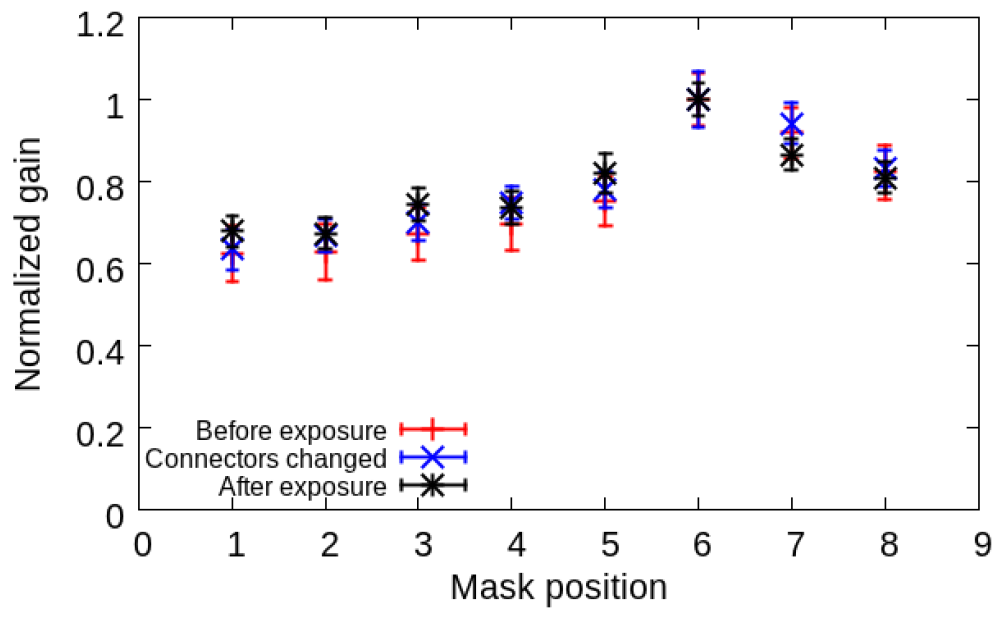} 
\caption{Gain vs position (before, during and after  irradiation period) for exposed (top) and non-exposed (bottom) detector. The exposure region is indicated.}
\label{gain_pos}
\end{figure}

In order to study the relative gain homogeneities before and after exposure, measurements on the two detectors has been performed at different positions before the ageing period, when the grounding connectors have been removed and after the exposure. The relative gain at each position for these three set of measurements is plotted in Fig. \ref{gain_pos} and shows that the gain profile in both detectors keeps the same structure, and is compatible with previous measurements, at different aging periods. Moreover, the exposed detector region do not show a relevant different in relative gain compared to the other non-exposed regions. The X-rays irradiation has no effect on the detector behaviour.

More details about this study can be found in \cite{javier}

\section{Conclusion}

The different anode-resistive technologies that have been tested show very good performances, with a good spatial resolution (better than 100 $\mu$m for a pitch of 0.5 mm), a good efficiency (better than 95\%) and an efficient sparks reduction.\\

The R-strip-to-ground technology (R17) seems to be the best candidate for the HL-LHC environment since it presents a better efficiency, a very good spark reduction (no voltage drop -- no dead time), a good spatial resolution achievable with a pitch of 0.5 mm, a better robustness (compared to R-coating) and no charging effect (compared to Carbon-Loaded-Kapton).\\

A 2-Dimensional readout upgrade of this technology has been exposed during more than 21 days to an X-rays irradiation. The total integrated charge is almost 1C and corresponds to 5 years of HL-LHC with a security factor more than 5. After exposure, non ageing effect was visible.\\

A thermal neutron irradiation is already scheduled during November 2011 near the OrphŽe reactor at Saclay and a large area  resistive-anode chamber built at CERN workshop has been tested in July  2011 \cite{joerg}.

\section*{Acknowledgment}
The 2-dimensional anode-resistive detectors used for the ageing study have been provided by R. de Oliveira and J.Wotschack from CERN.

\end{document}